\documentstyle{lamuphys}
\makeatletter
\let\chapter\hid@chapter
\makeatother
\input epsf
\begin{document}
\def\refer{\par\noindent\hangindent 20pt}
\pagenumbering{arabic}
\title{Search for Galaxies at $z>4$ from a Deep Multicolor Survey}

\author{E.\,Giallongo\inst{1}, S.\,Charlot\inst{2}, S.\,Cristiani\inst{3},
S.\,D'Odorico\inst{4}, A.\,Fontana\inst{1}}

\institute{Osservatorio Astronomico di Roma, I-00040 Monteporzio, Italy\\
giallo@coma.mporzio.astro.it
\and
Institut d'Astrophysique du CNRS, 98 bis Boulevard Arago, 75014 Paris, France 
\and
Dipartimento di Astronomia, Universit\`a di Padova, Vicolo dell'Osservatorio 
5,\\
I-35122 Padova, Italy
\and
European Southern Observatory, K. Schwarzschild Strasse 2, D-8046 Garching,\\
Germany
}

\maketitle

\begin{abstract}
We present deep BVrI multicolor photometry in the field of the quasar
BR1202-07 ($z_{em}=4.694$) aimed at selecting field galaxies at $z>4$.
We compare the observed colors of the galaxies in the field with those
predicted by spectral synthesis models including UV absorption by the
intergalactic medium and we define a robust multicolor selection of
galaxies at $z>4$. We provide spectroscopic confirmation of the high
redshift QSO-companion galaxy ($z=4.702$) selected by our method.
The first estimate of the surface density of galaxies in the redshift
interval $4<z<4.5$ is obtained for the same field, corresponding to a
comoving volume density of $\sim 10^{-3}$ Mpc$^{-3}$. This provides a
lower limit to the average star formation rate of the order of
$10^{-2}$ M$_{\odot}$ yr$^{-1}$ Mpc$^{-3}$ at $z\sim 4.25$.
\end{abstract}
\section{Introduction}

Deep images from the Hubble Space Telescope and ground based
telescopes (Keck, NTT, CFHT) are providing new exciting information
about abundance and morphology of galaxies in a wide redshift interval
up to $z\sim 4.5$.

In particular the search of high redshift galaxies is relevant not
only to extract information about the physical processes which control
the formation of individual objects, but also to probe the cosmological
evolution of the formation of galactic structures in the Universe.
Indeed cosmological scenarios are attempting to follow the evolution
in cosmic time of the galaxy formation, describing in detail the history
of the star formation. In the standard CDM cosmology, for example,
most of the stars are formed at intermediate redshifts ($z\sim 1$;
e.g. Cole et al. 1994).

A useful parameter which allows a more direct comparison between
theoretical predictions and data interpretations is the star formation
rate per unit comoving volume. A reference value at the present epoch
of $\sim 5\times 10^{-3}$ M$_{\odot}$ yr$^{-1}$ Mpc$^{-3}$ (for a
Salpeter IMF) has been recently given by Gallego et al. (1995) on the
basis of an H$\alpha$ galaxy survey.

Recent estimates of the galaxy luminosity function of faint galaxies
up to $z=1$ are providing the first evidence of strong evolution by a
factor of ten of the cosmological star formation rate in the redshift
interval $z=0-1$ (Lilly et al. 1996; Cowie et al. 1996).  This of
course implies that more than half of the stars formed at
intermediate redshifts, in good agreement with theoretical
expectations.

Nevertheless, the same models predict a fraction $< 2$\% of the
present mass density in stars at $z>4$ (Cole et al. 1994). It is
therefore at these very high redshifts that cosmological scenarios for
galaxy formation are more vulnerable to observational constraints.

Efficient selection criteria are needed to find out galactic
structures at these very high redshifts. Well known examples of high
$z$ sources are luminous Active Galactic Nuclei like quasars. Their
absorption spectra provide unique information on the abundance and
ionization state of the intergalactic medium at very high $z$. The
presence of the IGM has a twofold cosmological relevance. First, Lyman
absorption by the IGM along any line-of-sight produces strong
depression of the UV spectrum of high redshift sources.  Moreover, its
high ionization level requires a large background of UV ionizing
photons up to $z\sim 4-5$ (Giallongo et al. 1994,1996). This large UV
background is only marginally consistent with that produced by the
observed quasars (Haardt \& Madau 1996), leaving room for a possible
UV contribution by a large number of star-forming galaxies at $z>4$.

\section{The Multicolor Selection}

In selecting galaxies which are actively forming stars at very high
redshifts, two different approachs can be followed. It is possible to
exploit the intrinsic spectral properties expected from star formation
activity, or, in case we want to select galaxies at $z>4$, it is
better to exploit the complex but universal opacity to the UV photons
of the intergalactic medium.

The selection criteria based on the intrinsic spectral properties
exploit the main UV features of the star-forming galaxies like the
possible presence of strong emission lines and/or the Lyman absorption
break of the flat UV continuum due to the stellar evolutionary
properties plus Lyman continuum absorption by the interstellar medium
present inside the same galaxy.

Surveys based on the detection of intense Lyman alpha emission by
means of narrow band imaging in the optical/IR band in the redshift
interval $1.8<z<6$ have provided no systematic detections of high $z$
galaxies with only few exceptions (e.g. Macchetto et al. 1993).

A very efficient method based on the detection of the Lyman break
present in a flat rest-frame UV continuum has been proposed by Steidel
\& Hamilton (1992). An appropriate choice of a set of broad band
colors can allow the detection of the Lyman break in a given,
relatively narrow redshift interval. Steidel used in particular a set
of U G R filters adopted to select Lyman break galaxies in the
redshift interval $2.8<z<3.4$.  Since the spectrum of an actively
star-forming galaxy is flat longward of the Lyman break, an average
color of G--R$\sim 0.5$ is expected in the selected redshift
interval. At the same time strong reddening is expected in the U--G
color which samples the drop of the emission shortward of the Lyman
break (U--G$>1.5$) for $z\sim 3$ galaxies.

Given the faintness of the galaxies ($R\sim 24-25$), low resolution
spectra with good s/n are beeing produced only in the last period from
observations with the Keck LRIS instrument. Steidel et al. (1996) 
confirmed with low
resolution spectra the identification of 15 galaxy candidates in the
expected redshift interval showing the high success rate
($>70$\%) of this multicolor selection. Extrapolating the success rate
obtained for the subsample of their candidates, Steidel et al. (1996)
provide a first estimate of the surface density of galaxies at $z\sim
3$ of the order of 0.4 arcmin$^{-2}$ corresponding to a comoving
volume density of $3.6\times 10^{-4}$ h$_{50}^3$ Mpc$^{-3}$. The
average rest frame UV luminosity of these galaxies would imply a
cosmological star formation rate SFR$\sim 3\times 10^{-3}$ M$_{\odot}$
yr$^{-1}$ Mpc$^{-3}$.

However, the selection of galaxies at redshift $z>4$ becomes
considerably more efficient if we take into account the complex
absorption produced by the intergalactic medium in the UV spectrum of
high $z$ sources. The reddening produced by the IGM in the colors of
high $z$ quasars was investigated by Giallongo \& Trevese (1990) and a
considerable number of very high $z$ quasars has been discovered by
means of this multicolor technique (e.g. Warren et al. 1991; Irwin,
McMahon \& Hazard 1991). Recently, Madau (1995) has refined and
applied this multicolor method to the selection of high $z$ star
forming galaxies.

\vskip 6.5truecm

{~ 
\epsfxsize=250 pt 
\includegraphics{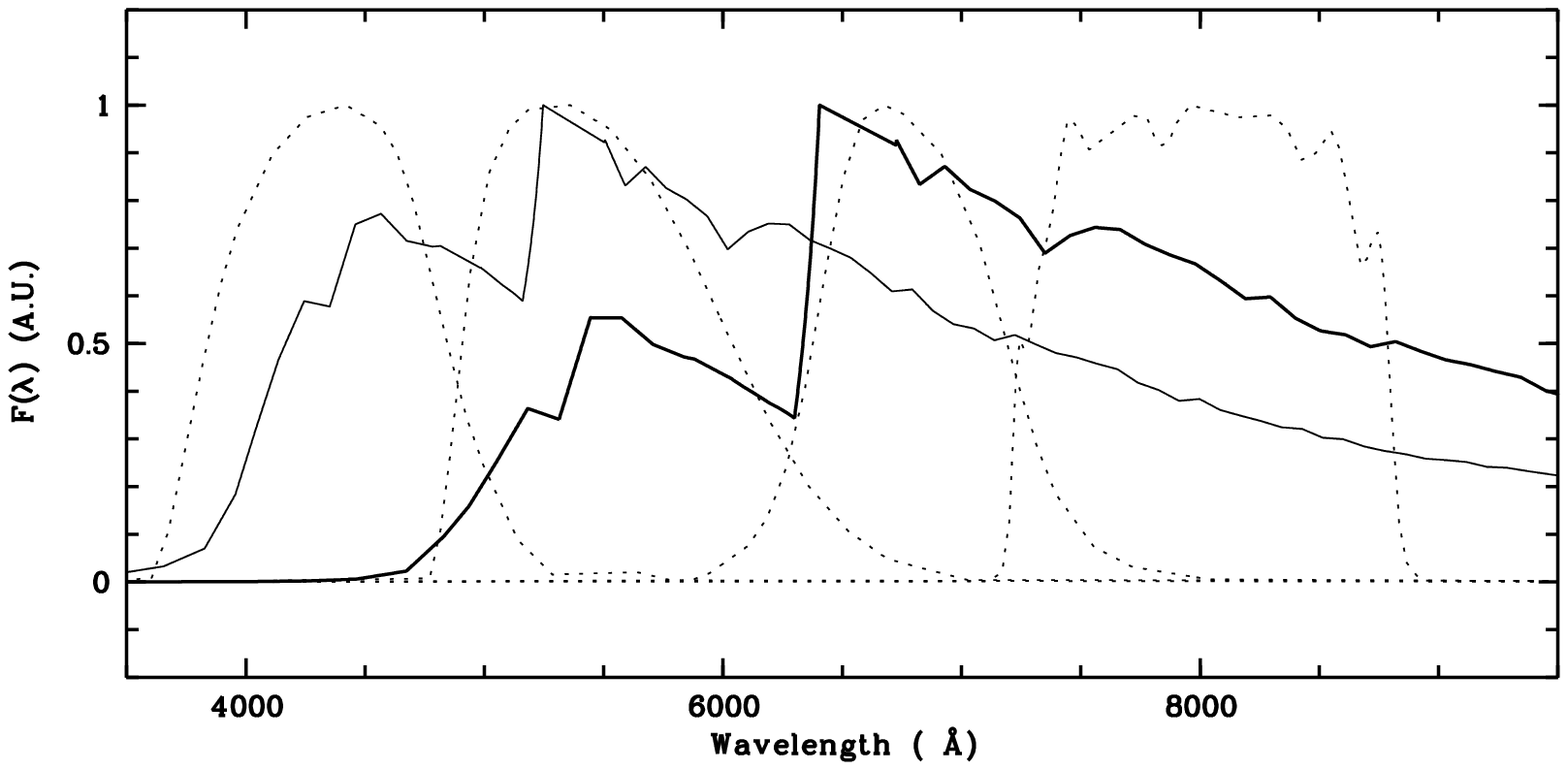} 

\noindent
{\bf Fig. 1}. Spectra of constant star-forming galaxy emitting at $z=3.25$
(thin line) and $z=4.25$ (thick line) depressed by Lyman absorption by the
intergalactic medium. A BVrI filter set (from left to right) has been
superimposed (dashed lines).\\

We have plotted in Fig.1 the average IGM absorption affecting the
spectral properties of a constant star-forming galaxy emitting at
$z=3.25$ or $z=4.25$. We have adopted the Madau (1995) absorption model
and the galaxy spectrum by Bruzual \& Charlot (1993) with a Salpeter
IMF.  First, it is to notice that, at given redshift, the absorption
by IGM is characterized by the average Lyman alpha forest absorption
present just shortward of the galaxy Lyman alpha wavelength and by the
absorption of the overall Lyman series down to the Lyman continuum
absorption, where the IGM is fully opaque to the UV radiation. While
at $z\sim 3$ the Ly$\alpha$ forest absorption produces a fractional
decrement of only $\sim 30$\%, at $z\sim 4.5$, 60--70\% of the galaxy
emission is lost causing a strong and easily detectable reddening in
the broad band colors which sample the relevant wavelength interval.

An efficient sampling of this complex absorption requires at least 4
broad band filters. We have chosen the BVI Johnson and Gunn r filters
to extend the multicolor selection up to $z\sim 4.5$ (Fontana et
al. 1996). These filters are plotted in Fig.1 superimposed to the
galaxy spectra.

The r-I color can select the intrinsic flat spectrum of any
star-forming galaxy up to $z\sim 4.5$, while the V-r and B-r colors
provide evidence of the strong reddening expected because of the
Ly$\alpha$ and Lyman continuum IGM absorption, respectively.

To examine how robust is the color selection of $z>2$ galaxies, we
have computed the expected colors as a function of redshift in our
photometric system (Fontana et al. 1996) adopting the spectral
synthesis Bruzual \& Charlot (1993) model. Models of these kind have a
number of parameters whose uncertainties can be large in some
cases. However, the resulting color changes of a few tenths do not
alter the robustness of our color selection, as shown in the
following.

To explore how the colors of different galaxy spectral populations are
distributed in redshift, we have considered the {\it e-folding}
star-formation timescale $\tau$ as the main interesting
parameter. Different $\tau$ values reproduce different spectral
types. For example, a star-formation timescale of $\tau\sim 1$ Gyr is
more appropriate for an early type galaxy, while $\tau >3$ Gyr
represent the spectral properties of different late type galaxies.  At
each ``observed'' redshift, different ages (i.e. different formation
redshifts ranging from 1 to 7) have been considered for galaxies with
a given $\tau$. A Salpeter IMF and a solar metallicity have been
adopted.

Our relevant colors B-r, V-r, r-I are reproduced as a function of $z$
in Fig.2 only for the case $\tau =1$ Gyr.

The first remark that should be done is that the r-I color is sampling
the intrinsic spectrum of galaxies in a wide redshift interval from
$z=0$ to $z\sim 4.5$. At $z>4.5$ IGM absorption in the r band produces
appreciable reddening in the r-I colors. In selecting galaxies in the
redshift range $2.5<z<4.2$ the fundamental property of galaxies of all
spectral types is the flatness of their rest-frame UV spectra revealed
in their r-I colors (see Fig.2). Indeed, in the relevant $z$
interval is always r-I$<$0.2 due to the intense star formation
activity. At $z<2.5$ the r-I colors are sampling progressively longer
rest-frame wavelengths where the galaxy spectra are in general
steeper, always resulting in r-I$>$0.2. Thus, it appears that the r-I
color selection is very useful to discriminate high $z$
galaxies in the field. Of course the presence of non-negligible
photometric errors suggests the use of bluer colors to select high $z$
galaxies with high confidence.

From Fig.2 it can be seen that the IGM absorption produces strong
reddening first in the B-r colors with B-r$\sim$1 at $z\sim 3$ then in
the V-r color with V-r$\sim$1 at $z\sim 4$. Thus, the simultaneous
presence of the three colors at the average expected values can select
galaxies at $\langle z\rangle \sim 3$ and at $\langle z\rangle \sim 4$
or even more. Any possible contamination by an old population with
steep blue spectra (a pronounced 4000 {\AA} break) producing red B-r
and V-r colors at $z=0.5-1$ can be avoided just requiring a ``flat''
r-I color.

\vskip 9truecm

{~
\epsfxsize=250 pt
\includegraphics{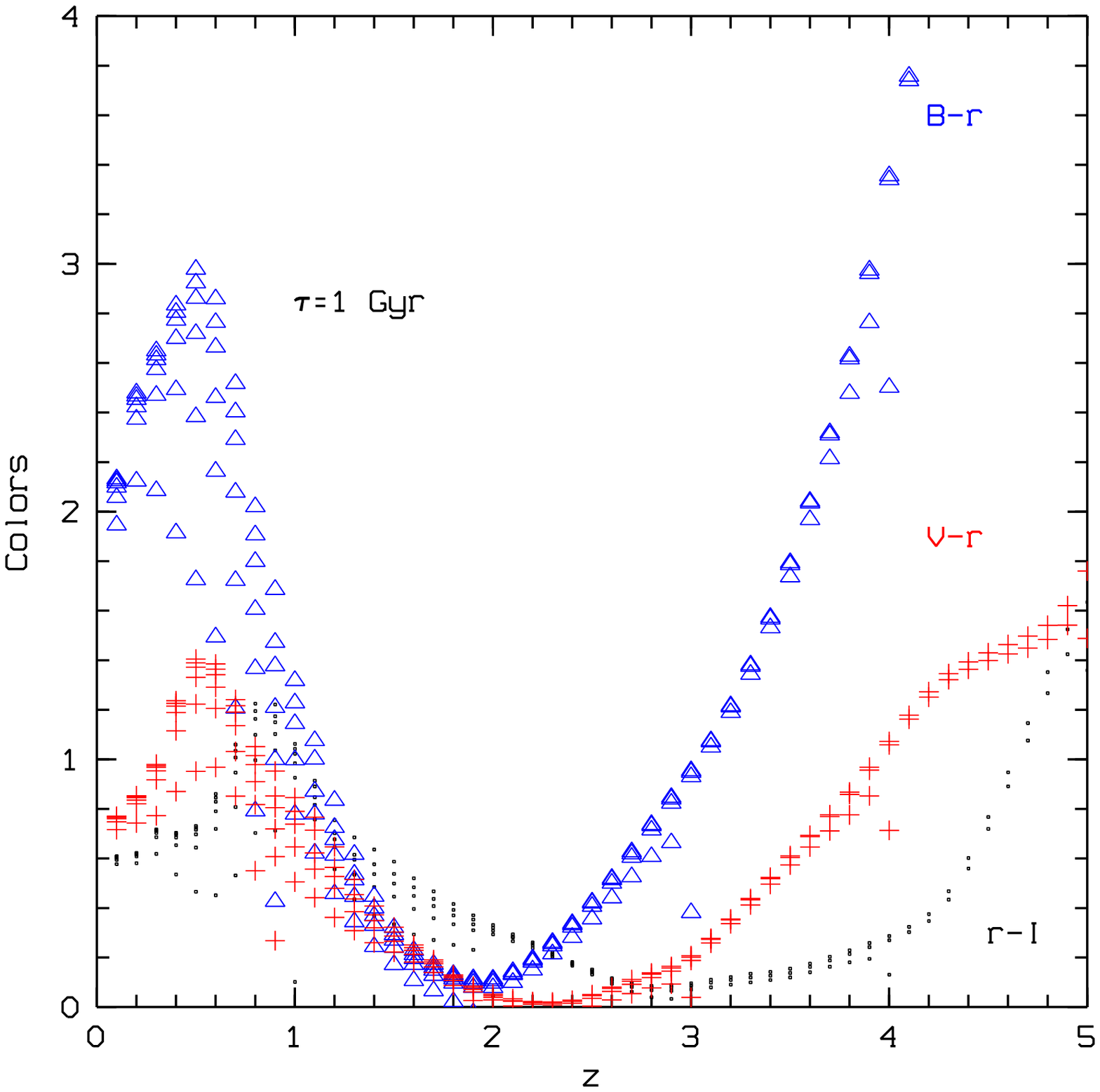}
}

\noindent
{\bf Fig. 2}. Colors as a function of redshift for galaxies with 
star-formation timescale $\tau = 1$ Gyr. Different formation redshifts
have been adopted in the interval $z=1-7$.

\section{A QSO Companion Galaxy at $z=4.702$}

We have applied this multicolor technique to the field around one of
the brightest high $z$ QSO BR1202-07 at $z=4.694$ (McMahon et
al. 1994, Storrie-Lombardi et al. 1996) where at
least one very high $z$ galaxy is close to the line of sight to the QSO
as shown by the detection of a damped absorption system at
$z\sim 4.4$ (Giallongo et al. 1994).

Deep BVrI images were obtained during the 1994 at the NTT with the
SUSI direct imaging CCD camera in very good seeing conditions
(FWHM$\sim$ 0.5-0.6 for the stellar objects in the r and I images). A
diffuse object clearly stands out 2.2 arcsec NW of the QSO with an r
magnitude of r=24.3. The companion galaxy has the unusual colors
expected for star-forming galaxies at $z>4$, i.e.  r-I=0.2, V-r=1.9,
B-r$>$3 (Fontana et al. 1996). On the basis of our multicolor
selection criterion we estimated a probable redshift range $4.4<z<4.7$
depending on the intensity of the galaxy Lyman alpha emission. On the
basis of the 1500 {\AA} continuum flux measured in the I band we
derived a star formation rate $\sim 16$ M$_{\odot}$ yr$^{-1}$ (for a
Salpeter IMF). This galaxy has also been detected in the K band by
Djorgovski who estimated a magnitude K$\sim 23$. The r-K$\sim$1 color
so derived implies a very young age $<10^8$ yr independently of
details on the assumed metallicity and star-formation timescale.

This galaxy has also been observed in narrow band imaging centered at
the Lyman $\alpha$ QSO redshift by Hu et al. (1996) and in imaging
spectroscopy by Petitjean et al. (1996). Both authors discovered a
Lyman $\alpha$ emission in the galaxy spectrum at $z\sim 4.7$.  The
strong Ly$\alpha$ emission increases the r flux ($\Delta$r$\sim -0.8$
mag) keeping a flat r-I color up to $z\simeq 4.7$ despite the strong
attenuation in the r band due to the presence of the Ly$\alpha$
forest.

We have recently obtained a low resolution (15 {\AA}) spectrum of this
galaxy at the NTT with EMMI (D'Odorico et al. 1996, in preparation)
which extends well in the red up to 9000 {\AA}. The spectrum is shown
in Fig.3 where the strong Ly$\alpha$ emission is detected at $z=4.702$
corresponding to a proper distance from the QSO $\sim 600$ kpc or
equivalently to a velocity difference $\Delta v\sim 400$ km s$^{-1}$.
The line flux $f\simeq 2\times 10^{-16}$ ergs s$^{-1}$ cm$^{-2}$
corresponds to a luminosity $L_{Ly{\alpha}}\simeq 3.8\times 10^{43}$
ergs s$^{-1}$. Although this line luminosity could be formally
converted into a star-formation rate, some contamination by
reprocessing of the QSO UV continuum can not be excluded even at
distances $\sim 100$ kpc. More important is the absence of any CIV
emission within the flux measured in the I band. This implies that the
redshifted I flux can be converted in the star-formation rate of $\sim
16$ M$_{\odot}$ yr$^{-1}$ previously mentioned.

\vskip 6.5truecm

{~ 
\epsfxsize=250 pt 
\includegraphics{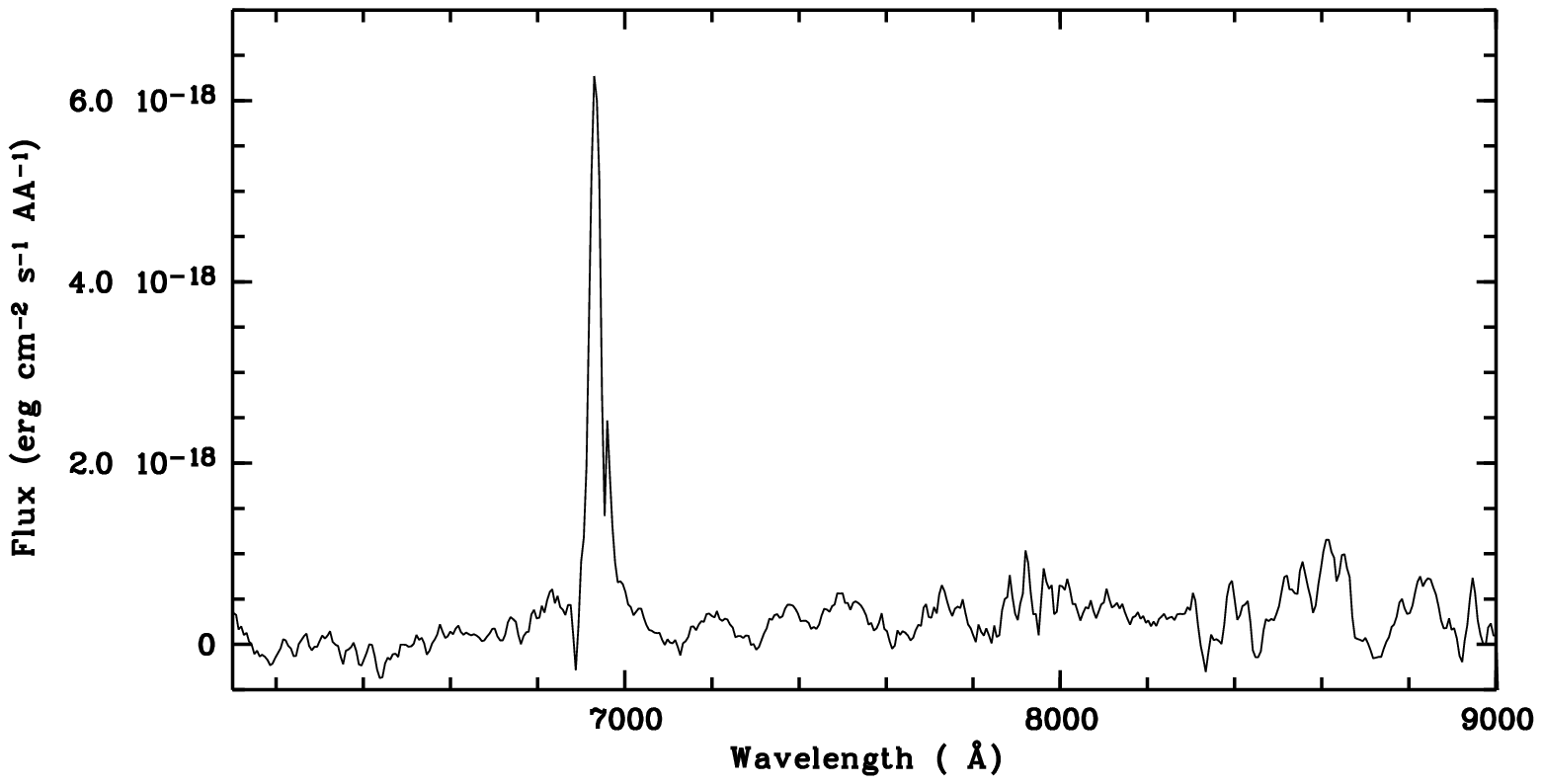} 
}

\noindent
{\bf Fig. 3}. NTT 15 {\AA} spectrum of the QSO-galaxy companion
showing a Ly$\alpha$ emission at $z=4.702$.\\

These observations provide the first evidence of strong star formation
activity at $z\ga 4.5$.

\section{A Sample of High Redshift Galaxies}

In the 2.2$\times 2.2$ arcmin$^2$ field centered on the QSO position
we have detected and counted galaxies in the r band down to r$\simeq
26$ mag by means of the SExtractor software package (Bertin 1994).
Reliable colors have been obtained for galaxies with r$\leq 25$ mag.
We have selected galaxies in two different redshift ranges. First,
galaxies satisfying the criterion r-I$<$0.2 and B-r$>$1 are expected
to lie in the redshift interval $3\la z \la 4$. We found 11 galaxies
at r$\leq 25$ mag in this $z$ interval corresponding to a surface
density of 2.3 arcmin$^{-2}$. The derived average comoving volume
density at $\langle z \rangle \simeq 3.5$ is $\phi \sim 10^{-3}$
Mpc$^{-3}$. The redshift interval $4\la z\la 4.5$ has been selected
imposing r-I$<$0.4, V-r$>$1 and B-r$>$2. We found 5 galaxies in the
field corresponding to a surface density of 1 arcmin$^{-2}$ and to a
comoving volume density $\phi \sim 10^{-3}$ Mpc$^{-3}$ at $\langle z
\rangle \simeq 4.25$ (see Fig.4).

\vskip 6.5truecm

{~ 
\includegraphics{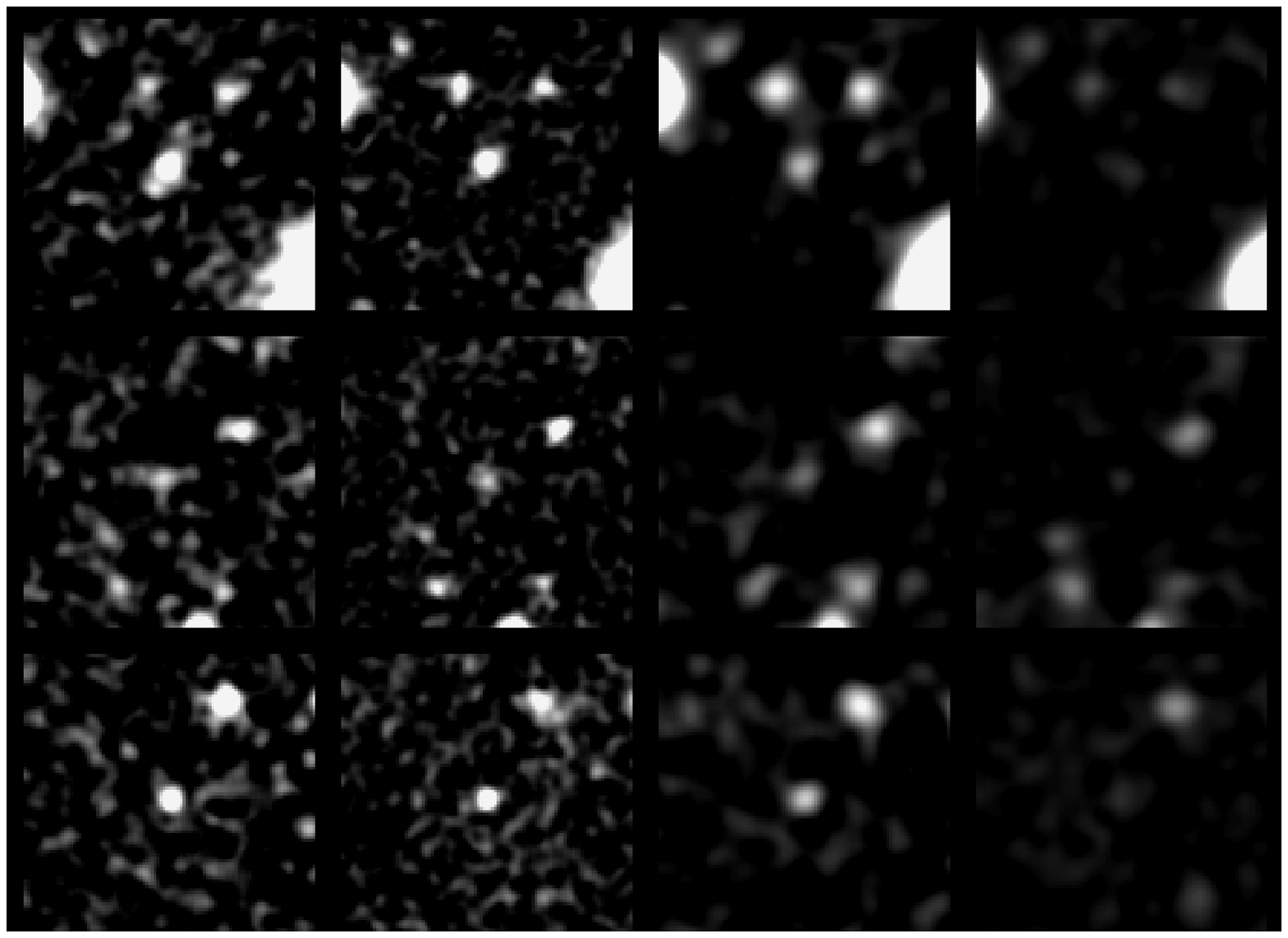} 
}

\noindent
{\bf Fig. 4} IrVB (from left to right) images of three $z\ga 4$ galaxy
candidates.\\

Of course these estimates have to be considered as lower limits since
galaxies at fainter r magnitudes will contribute somewhat to the
volume density. Moreover, the selected galaxies have colors consistent
with dust free spectral models. Although an intrinsic reddening
E(B-V)$\la 0.1$ does not alter appreciably the r-I color selection,
some high $z$ dusty galaxies could be lost by our multicolor
selection.
The average $\langle I \rangle \sim 24.5$ mag of the galaxies at $3\la
z\la 4.5$ corresponds to an average star-formation rate $\sim 8$
M$_{\odot}$ yr$^{-1}$.  The corresponding cosmological SFR per unit
comoving volume is $\sim 10^{-2}$ M$_{\odot}$ yr$^{-1}$ Mpc$^{-3}$ in
agreement with the value found by Steidel et al. (1996) at $\langle z
\rangle \simeq 3.25$.  This limit is about 2 times higher than the
present value derived by Gallego et al. (1995) assuming a Salpeter IMF
and 5 times lower that at $z\sim 1$ (Lilly et al. 1996). Thus the
cosmological SFR increases by a factor of 10 from $z=0$ to $z=1$ then
it seems to decline by a factor 5 or less up to $z=4.5$. Assuming a
fiducial local stellar mass density $\sim 3\times 10^8$ M$_{\odot}$
Mpc$^{-3}$ (Cowie et al. 1995) and an age for the $z>4$ galaxies of a
few $10^8$ yr, a lower limit to the luminous matter density at $z\sim
4.25$ could be of the order of 1\% of the local value. Of course our
estimates are derived in a small field of 4.8 arcmin$^2$ centered on a
high $z$ QSO. Larger areas are needed to reduce density fluctuations.

\section{Prospects for the VLT}

The large collecting area of the VLT can be exploited to confirm and
study high $z$ galaxy candidates selected by multicolor photometry.
The first and most obvious follow-up is the spectroscopic observation
of galaxies down to r$\sim 25$ mag by means of the mos capability
present in FORS. For objects fainter than r$\sim 25.5$, a different
approach has to be pursued. Intermediate band filters (200--300 {\AA})
can be used to extend the redshift identification to r$\sim 26.5-27$
mag in a reasonable observing time (Fontana et al. 1996, this volume).

%
%
%
%
%

\section{References}

\refer Bertin,\, E. (1994), SExtractor Manual
(Paris: IAP)

\refer Bruzual,\, A. G., Charlot,\, S.,
(1993): ApJ, 405, 538

\refer Cole,\, E., Arag\'on-Salamanca,\, A.,
Frenk,\, C. S., Navarro,\, J. F., Zepf,\, S. E. (1994): MNRAS, 271, 781

\refer  Cowie,\, L. L., Hu,\, E. M.,
Songaila,\, A. (1995): Nature, 377, 603

\refer Cowie,\, AL. L.,Songaila,\ A., Hu,\, E.
 M., Cohen,\, J. G. (1996) AJ in press

\refer Fontana,\, A.,
Cristiani,\, S., D'Odorico,\, S., Giallongo,\, E., Savaglio,\, S. (1996): 
MNRAS, 279, L27

\refer Gallego,\, J., Zamorano,\, J.,
Arag\'on-Salamanca,\, A., Rego,\, M. (1995): ApJ, 455, L1

\refer Giallongo,\, E., Cristiani,\, S.,
D'Odorico,\, S., Fontana,\, A., Savaglio,\, S. (1996): ApJ, in press

\refer Giallongo,\, E., D'Odorico,\, S.,
Fontana,\, A., Savaglio,\, S., McMahon,\, R. G., Cristiani,\, S.,
Molaro,\, P., Trevese,\, D. (1994): ApJ, 425, L1

\refer Giallongo,\, E., Trevese,\, D.,
(1990): ApJ, 353, 24

\refer Haardt,\, F., Madau,\, P.,
(1996): ApJ, 461, 470

\refer Hu,\, E. M., McMahon,\, R. G., Egami,\, E.
(1996): ApJ, 459, L53

\refer Irvin,\, M. J., McMahon,\, R. G.,
Hazard,\, C. (1991): {\it The Space Distribution of Quasars}
(San Francisco: ASP), 21, 117

\refer Lilly,\, S. J., Le F\'evre,\, O.,
Hammer,\, F., Crampton,\, D. (1996): ApJ, 460, L1

\refer Macchetto,\, F. D., Lipari,\, S., 
Giavalisco,\, M., Turnshek,\, D. A., Sparks,\, W. B. (1993): ApJ, 404, 511

\refer Madau,\, P. (1995): ApJ, 441, 18

\refer McMahon,\, R. G., Omont,\, A.,
Bergeron,\, J., Kreysa,\, E., Haslam,\, C. G. T. (1994): MNRAS, 267, L9

\refer Petitjean,\, P., P\'econtal,\, E.
Valls-Gabaud,\, D., Charlot,\, S. (1996): Nature, 380, 411

\refer Steidel,\, C. C., Giavalisco,\, M., 
Pettini,\, M., Dickinson,\, M., Adelberger,\, K. L. (1996): ApJ, 462, L17

\refer Steidel,\, C. C., Hamilton,\, D.,
(1992): AJ, 104, 941

\refer Storrie-Lombardi,\, L. J., 
McMahon,\, R. G., Irwin,\, M. J., Hazard,\, C.
(1996): ApJS, in press

\refer Warren,\, S. J., Hewett,\, P. C.,
Osmer,\, P. S. (1991): ApJS, 76, 23
\end{document}